\documentclass[lettersize,journal]{IEEEtran}
\usepackage{amsmath,amsfonts}
\usepackage{algorithmic}
\usepackage{array}
\usepackage[caption=false,font=normalsize,labelfont=sf,textfont=sf]{subfig}
\usepackage{textcomp}
\usepackage{stfloats}
\usepackage{url}
\usepackage{verbatim}
\usepackage{graphicx}
\usepackage{cite}
\newtheorem{theorem}{\normalfont \textbf{Theorem}}
\newtheorem{definition}{\normalfont \textbf{Definition}}

\newtheorem{corollary}{\normalfont \textbf{Corollary}}

\usepackage{mathtools}

\usepackage[ruled]{algorithm2e}
\usepackage[T1]{fontenc}
\usepackage{comment}
\usepackage{amsmath}
\usepackage[dvipsnames]{xcolor}
\usepackage{float}

\begin{document}

\title{An Incentive-Compatible Reward Sharing Mechanism for Mitigating Mirroring Attacks in Decentralized Data-Feed Systems}

\author{Sina Aeeneh,~
        Nikola~Zlatanov,~
        and~Jiangshan~Yu~
        
\thanks{Sina Aeeneh is with the Department of Electrical and Computer Systems Engineering, Monash University, Clayton, VIC 3800, Australia\protect\\ E-mail: sina.aeeneh@monash.edu}
\thanks{Nikola Zlatanov is with the Department of Computer Science and Engineering, Innopolis University, Innopolis, Respublika Tatarstan, Russia, 420500.}
\thanks{Jiangshan Yu is with the School of Computer Science, The University of Sydney, Camperdown, NSW 2006, Australia.}
}

\markboth{}%
{Shell \MakeLowercase{\textit{Aeeneh et al.}}:An Incentive-Compatible Reward Sharing Mechanism for Mitigating Mirroring Attacks in Decentralized Data-Feed Systems}


\maketitle

\begin{abstract}
Decentralized data-feed systems enable blockchain-based smart contracts to access off-chain information by aggregating values from multiple oracles. To improve accuracy, these systems typically use an aggregation function, such as majority voting, to consolidate the inputs they receive from oracles and make a decision. Depending on the final decision and the values reported by the oracles, the participating oracles are compensated through shared rewards. However, such incentive mechanisms are vulnerable to \textit{mirroring attacks}, where a single user controls multiple oracles to bias the decision of the aggregation function and maximize rewards. This paper analyzes the impact of mirroring attacks on the reliability and dependability of majority voting-based data-feed systems. We demonstrate how existing incentive mechanisms can unintentionally encourage rational users to implement such attacks. To address this, we propose a new incentive mechanism that discourages Sybil behavior. We prove that the proposed mechanism leads to a Nash Equilibrium in which each user operates only one oracle. Finally, we discuss the practical implementation of the proposed incentive mechanism and provide numerical examples to demonstrate its effectiveness.
\end{abstract}


\section{Introduction}

\IEEEPARstart{D}{ecentralized} smart contracts are pivotal elements of blockchain systems, offering the potential to revolutionize various industry sectors \cite{8494045, khan2021blockchain}. Smart contracts are essentially instructions or agreements written as pieces of code stored on a blockchain. Despite their huge potential to bring automation, transparency, and reliability to business operations, their practical implementation faces challenges in applications where the execution of the smart contract depends on off-chain information about real-world events, which is obtained from off-chain sources such as sensors, imaging satellites, and human users.

For example, consider a city council outsourcing the cleaning, repair, and maintenance of streets, parks, and other public spaces to an external contractor. Instead of signing a conventional paper-based contract, the two parties may agree to publish a smart contract on a blockchain platform to enhance transparency and facilitate automation. When maintenance issues arise, or a facility needs repairing, residents can report them through blockchain transactions. If the contractor fails to resolve the problem promptly, penalties may apply according to the smart contract terms. This can be done automatically by deducting funds from the contractor's cryptocurrency account and transferring them to the city council account. However, to validate such incidents and execute the smart contract's terms, the blockchain network in this case would require off-chain data in the form of evidence and testimonies from independent third parties that can serve to make a decision whether the contract between the city council and the contractor has been violated.

Since blockchains are isolated systems, an intermediary component, commonly known as a data-feed system, is needed to relay information from off-chain information sources to the blockchain network. Data-feed systems may utilize various tools and techniques, including financial incentive mechanisms, to ensure the integrity and reliability of the off-chain information that is provided to the blockchain.

Data-feed systems can be configured in either a centralized or decentralized manner~\cite{Oracle-Review}. In a centralized setup, the parties to a smart contract explicitly determine their trusted data sources, and a single oracle is utilized to read the required information from these sources and send it to the data-feed system, which then makes a decision based on this information. In the above example, for instance, a supervising company could be appointed to investigate incidents and report their findings to the blockchain, acting as a trusted data source and a centralized data-feed system. To ensure that centralized data-feed systems cannot tamper with the information they receive from off-chain data sources, various solutions have been proposed, including a trusted hardware-based protocol \cite{town}, a modified version of the TLS (Transport Layer Security) protocol \cite{Ritzdorf2017TLSNNO}, and a cryptographic protocol based on zero-knowledge proofs \cite{DECO}.

Centralized data-feed systems are easy to implement, however, their reliance on a single oracle and a fixed set of information sources can introduce single points of failure, bringing all the problems associated with centralized systems back to the blockchain world \cite{Oracle-Review}. For instance, centralized data-feed systems can be targeted by denial-of-service or bribery attacks. To address these issues, several decentralized architectures have been proposed for data-feed systems~\cite{ASTRAEA, ChainLink-whitepaper, Witnet}. 

In a decentralized data-feed system, each off-chain data request is handled by multiple oracles in parallel, with each oracle connecting to a subset of the available information sources to retrieve the information. An aggregation function (AF) then receives the subjective values reported by the oracles, evaluates them, estimates the objective value of the parameter of interest, and makes a decision. In the above example, the local community can serve as independent oracles and submit their observations to the blockchain, possibly through a mobile application. These submissions can then be used to assess whether the contractor is meeting its obligations or not. Implementing a decentralized data-feed system eliminates single points of failure and enhances the system's resilience against bribery attacks.

ASTRAEA is a decentralized data-feed system that feeds binary information into blockchains~\cite{ASTRAEA}. Oracles in ASTRAEA can participate in either a low-risk/low-reward or high-risk/high-reward game. The final value of the data point is determined by the majority of the oracles playing the low-risk game and may be confirmed by the majority of the oracles playing the high-risk game. The authors in \cite{ASTRAEA} showed that the honest behavior of players constitutes a Nash Equilibrium strategy. However, the ASTRAEA protocol has limitations when it comes to broader applications involving non-binary data points \cite{SoKOracle}. Chainlink and Witnet are more adaptable decentralized data-feed systems that can be used for both binary and non-binary data \cite{ChainLink-whitepaper, Witnet}. In particular, in Chainlink, the AF is implemented in the form of a smart contract, known as the aggregator smart contract. The aggregator smart contract also incorporates an incentive mechanism for distributing tasks and rewards among oracles based on their reputation and performance. The detailed design and the analysis of the AFs and reputation systems are not discussed in Chainlink's original paper and, hence, require further research.  

Despite these developments, only a limited number of applications currently utilize decentralized data-feed systems for relaying off-chain information into blockchains. This can be attributed to a lack of comprehensive understanding of the reliability and dependability of such systems under various scenarios. In particular, making reliable decisions based on inputs from malicious oracles that strategically manipulate the aggregation function or from colluding oracles attempting to sway its outcome remains a fundamental challenge in decentralized systems \cite{Wang-TDSC-Strategic, Rezvani-TDSC-Colllusion}.

In this paper, we study the reliability of decentralized data-feed systems for retrieving off-chain data points that take on values from a limited set. We adopt the majority voting function (MVF), a natural candidate to implement as an AF in this setting \cite{ASTRAEA, ChainLink-whitepaper, Witnet, 576970}. 

The MVF's output is simply the value favored by the largest number of oracles. For binary data points, the authors in \cite{BlockchainOracleAdler} showed that the error rate of the MVF decays toward zero as the number of independent and identically distributed (i.i.d.) oracles with an error rate below $0.5$ increases. The work in \cite{ourpaper} extends \cite{BlockchainOracleAdler} to scenarios with non-binary data points and involving oracles that are independent but not necessarily identically distributed. More specifically, for general multi-class classification problems, \cite{ourpaper} provides necessary and sufficient conditions under which the error rate of the MVF decays exponentially toward zero as the number of mutually independent oracles increases. Conversely, if these conditions are not met, the error rate increases with the number of oracles. The results in \cite{ourpaper} show that accurate decentralized data-feed systems using an MFV as an AF are feasible when oracle responses are mutually independent.

Unfortunately, the existing incentive mechanisms of decentralized data-feed systems inadvertently encourage users to breach the mutual independence requirement through what is known as the \textit{mirroring attack} \cite{truthful}. This attack involves a user with access to off-chain data sources creating and controlling multiple oracles, all reporting the same data collected. Through implementing this attack, the user seeks to influence the outcome of the MVF-AF and attract more rewards. Since in an open and decentralized data-feed system, it is challenging for an external observer to link a real user to the oracles they own, it is practically impossible to detect the mirroring attack~\cite{ChainLink-whitepaper}.

Unlike blockchain consensus protocols, the mirroring attack is a major issue for decentralized data-feed systems. The main difference stems from the fact that in a decentralized data-feed system, the focus is on enhancing the accuracy of the provided off-chain information through the utilization of multiple oracles in parallel. In contrast, blockchain consensus protocols are designed to enhance the internal consistency, availability, and scalability of blockchains \cite{2021chainlink}, and therefore can rely on a single randomly chosen validator to generate each block. Previous studies have highlighted the threat that mirroring attacks pose to the reliability and dependability of decentralized data-feed systems~\cite{ChainLink-whitepaper, OracleProblem, SurveyOracle, 9154123, truthful}. However, most existing incentive mechanisms for blockchain-based information retrieval systems are either designed for centralized data-feed architectures, where mirroring attacks are not relevant~\cite{lin2023blockchain, yin2020blockchain, vakilinia2023incentive}, or assume threat models in decentralized settings that do not consider such attacks~\cite{9321132, hu2020blockchain}. To the best of our knowledge, only one paper has offered a high-level discussion about potential directions for addressing this problem \cite{truthful}. The authors in \cite{truthful} proposed a solution that involves reducing the growth rate of voting power while increasing the growth rate of the reward factor as an oracle's stake increases. Despite its high-level nature and lack of a detailed methodology for determining exact system parameters, the solution discussed in \cite{truthful} offers lower accuracy compared to the solution proposed in the current paper, which focuses solely on modifying incentives without altering voting powers. This is because adjusting the voting power of oracles merely based on their stakes moves the system away from the level of accuracy achievable through the optimal distribution of voting power needed to exploit the wisdom of the crowd.

In this paper, we introduce an algorithm that alters the incentive mechanism (IM) of decentralized data-feed systems to discourage users from executing the mirroring attack. We provide a game-theoretic analysis and outline the practical implementation of this revised IM. We believe that our proposed IM, promoting the mutual independence of oracles, coupled with recent findings on the reliability of the MVF for mutually independent oracles in \cite{ourpaper}, can create opportunities for the practical implementation of decentralized data-feed systems.

Our main contributions in this paper are listed as follows:
\begin{itemize}
    \item In Section \ref{sec:system model}, we introduce a mathematical model of decentralized data-feed systems and discuss the necessary conditions for a reliable data-feed system employing the MVF as an AF.  
    \item In Section 3, we discuss the mirroring attack and propose an IM to discourage users from carrying out this strategy. We show that with the proposed IM, the strategy where all users refrain from employing the mirroring attack forms a Nash Equilibrium, ensuring that the necessary conditions for a reliable data-feed system are satisfied.
    \item In Section 4, we discuss the practical implementation of the proposed IM and introduce an algorithm to determine its parameters.
    \item Finally, in Section 5, we provide numerical examples that demonstrate the effectiveness of the proposed algorithm in discouraging the mirroring attack, thereby enhancing the accuracy and reliability of decentralized data-feed systems.
\end{itemize} 

\paragraph*{Notation} The set \(\{1, 2, \dots , N\}\) is denoted by \([N]\). The notation \([N] \setminus m\) is used to denote \(\{n|\,n \in [N]  \text{ and }  n\neq m\}\). We use uppercase symbols, such as \(X\), to denote random variables and lowercase symbols, such as \(x\), for their realizations. Moreover, we use the calligraphic font to denote sets, e.g. \(\mathcal{X}=\{X_1, X_2,\dots, X_N\}\), and the bold font to denote vectors and matrices, e.g., \(\mathbf{X}=\left(X_1, X_2,\dots, X_N\right)\). Given two random variables, \(A\) and \(B\), the probability of \(A=a\) and probability of \(A=a\) given that \(B=b\) is denoted by \(\textrm{Pr}\left(A=a\right)\) and \(\textrm{Pr}\left(A=a|B=b\right)\), respectively. The notation \(|.|\) is used for denoting both the absolute value of variables and the number of elements in sets. Notations \(\lfloor{.}\rfloor\) and \(\lceil{.}\rceil\) denote the flooring and ceiling functions, respectively.
Finally, the indicator function is denoted by \(\mathbf{1}\left(.\right)\).

\section{System Model} \label{sec:system model}
In this section, we present a mathematical model of decentralized data-feed systems. 

Generally, a decentralized data-feed system consists of four types of components:
\begin{itemize}
    \item Data Source: Any entity such as remote sensors, imaging satellites, web pages, or human users that can supply information about real-world events. 
    \item Oracle: An agent that retrieves data from data sources and transmits it to the blockchain network in a standardized format. 
    \item Aggregation Function (AF): A deterministic function responsible for deriving the ultimate output of the data-feed system based on the values reported by the oracles.
    \item Incentive Mechanism (IM): A mechanism that distributes rewards among oracles according to a pre-determined rule, designed to compensate oracles for their work and incentivize them to provide accurate data in a timely manner.
\end{itemize}

In practice, the AF and IM can be combined within a single smart contract, which we refer to as the \textit{data-feed smart contract}. When a smart contract needs off-chain data to execute its terms, it sends a transaction to the data-feed smart contract. This transaction contains details about the requested off-chain data and a Service Level Agreement (SLA). The SLA sets performance expectations for data availability, accuracy, and latency, while also determining the total reward that would be distributed to the oracles for fulfilling the data query task, denoted by \(R\). Without loss of generality, we assume that \(R=1\) throughout this paper. 

Let \(X\) represent the data point that the smart contract tries to retrieve using the decentralized data-feed system. We model \(X\) as a discrete random variable with probability mass function (PMF) \(p\left(x\right)\). Throughout this paper, we assume that \(X\) takes values from the finite set \(\mathcal{X}=\left\{x_1,x_2, \dots, x_K\right\}\), leaving the continuous case for future works. Although the data provided through the oracles may be noisy and inaccurate, we assume that the users will not deliberately alter the off-chain information. However, we consider users to be rational and selfish, which could lead them to implement the mirroring attack to increase their average payoff. 

Let $N$ be the number of independent users, each controlling one or more oracles. We denote the \(n\)-th user by \(u_n\) and its set of oracles by \(\mathcal{O}_n^{(c_n)} = \left\{O_n^{(1)}, O_n^{(2)}, \dots, O_n^{(c_n)} \right\}\), where \(c_n\) is the number of oracles controlled by \(u_n\). In the special case where \(u_n\) has control over only one oracle, we use the notation $\mathcal{O}_n^*= \left\{O_n^{(1)}\right\}$ to denote its set of one oracle. The set of all oracles is represented as $\mathcal{O}=\bigcup_{n=1}^{N} \mathcal{O}_n^{(c_n)}$, while the set that encompasses all oracles, excluding those controlled by \(u_m\) is indicated by $\mathcal{O}_{-m}=\bigcup_{\underset{n\neq m}{n=1}}^{N} \mathcal{O}_n^{(c_n)}$.

Let \(Y_n^{(i)}\) denote the subjective opinion of the oracle \(O_n^{(i)}\) about the value of \(X\), where \(Y_n^{(i)} \in \mathcal{X}\) for all \(n \in [N]\) and \(i \in [c_n]\). Hence, \(Y_n^{(i)}\) is the value submitted by oracle \(O_n^{(i)}\) to the AF. In general, \(Y_n^{(i)}\) may differ from the true value \(X\) due to measurement noise or errors in data processing. The deviation of \(Y_n^{(i)}\) from \(X\) is modeled by a conditional PMF \(p\left(y_n^{(i)}\big|x\right)\). 

Since all oracles controlled by the same user experience the same observation noise for a given data point, their outputs follow the same conditional distribution. Therefore, we drop the index $i$ and denote the conditional PMF for any oracle $O_n^{(i)}$ in $\mathcal{O}_n^{\left(c_n\right)}$ as $p\left(y_n\big|x\right)$.

Let \(\mathbf{Y}\) denote the vector of values submitted by all oracles to the AF, given by
\begin{align} \label{eq:2100}
\mathbf{Y}=&\left( Y_1^{(1)}, Y_1^{(2)}, \dots, Y_1^{(c_1)}, Y_2^{(1)}, Y_2^{(2)}, \dots, Y_2^{(c_2)}, \right.\nonumber\\
    &\textrm{~~~~~~~~~~~~~~~~~~~~~~~~~~~~~~~~~~} \left. \dots, Y_N^{(1)}, Y_N^{(2)}, \dots, Y_N^{(c_N)}\right).
\end{align}
As the data-feed smart contract cannot distinguish between oracles based on their controlling users, the indices used in \eqref{eq:2100} are employed for explanatory convenience and do not precisely reflect how the AF receives the values from the oracles.

Once all oracles have submitted their values, or when the submission deadline specified in the SLA has elapsed, the AF generates the final output of the system, denoted by \(\hat{X}\), using the values in \(\mathbf{Y}\). The AF can be expressed mathematically as a function, \(f(\cdot)\), that maps \(\mathbf{Y}\) to \(\hat{X}\); i.e., 
\begin{equation}
f: \mathbf{Y} \to \hat{X},
\end{equation}
where \(\hat{X} \in \mathcal{X}\). The resulting value \(\hat{X}\) is then reported to the blockchain for use by smart contracts.

Finally, the IM distributes the total reward for the completed task among the oracles that have reported value that is deemed accurate, i.e., among the oracles that reported to the AF the value \(\hat{X}\). 

Our criterion for evaluating the performance of the data-feed system is the error rate defined as
\begin{equation}\label{eq:4}
    P_e= \textrm{Pr}\left(X\neq \hat{X}\right). 
\end{equation}
Hence, the optimal AF is an AF that minimizes \(P_e\) given by \eqref{eq:4}. For the worst case scenario in terms of minimising \(P_e\), we assume that \(X\) has a uniform distribution over its domain, i.e., \begin{equation}
    \textrm{Pr}\left(X=x_k\right)=\frac{1}{K}, \,\forall k \in \left[K\right]. 
\end{equation}
This assumption simplifies the equations throughout the paper, but also serves as the worst-case scenario in terms of error rate. With minor adjustments, our results can be extended to scenarios with non-uniform a-priori for $X$.

\begin{figure}[!htb]
    \centering
    \includegraphics[width=0.5\textwidth]{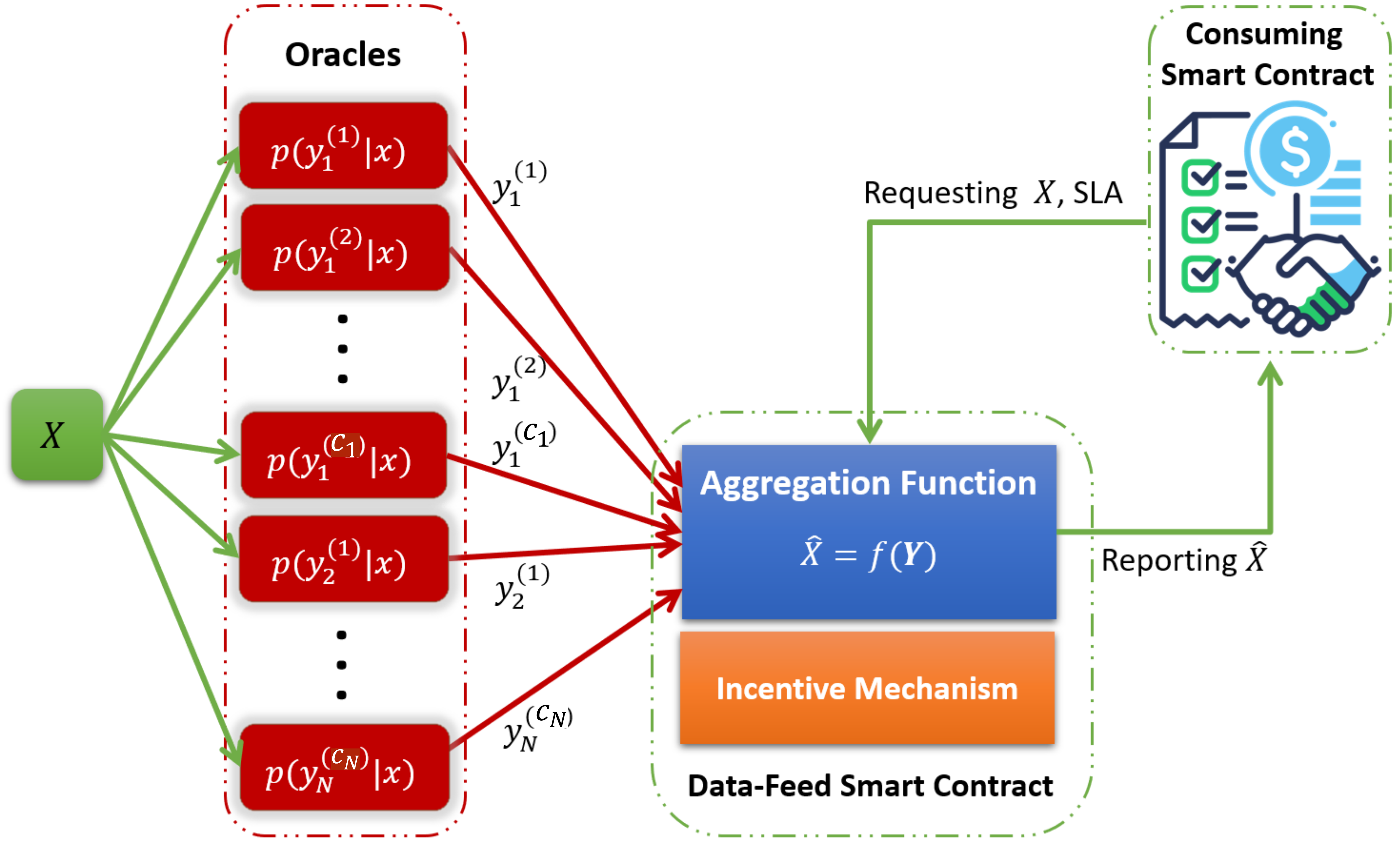}
    \caption{Sketch of the considered system model.}
    \label{fig:sketch}
\end{figure}

A sketch of the considered system is shown in Figure~\ref{fig:sketch}.

In the following subsections, we provide a detailed explanation and modeling of the AF and IM.

\subsection{The Aggregation Function}
In our system model, where the decentralized data-feed system retrieves off-chain information that takes value from a discrete set, the MVF is a suitable choice to implement as the AF. Therefore, in the rest of this paper, the AF \(f(\cdot)\) is defined as 
\begin{equation}\label{eq:7}
    f\left(\mathbf{Y}\right) 
    =\underset{x_k}{\textrm{argmax}} \sum_{n=1}^{N} \sum_{i=1}^{c_n} \mathbf{1}\left(Y_n^{(i)} = x_k\right).
\end{equation}
In other words, the AF-MVF outputs the class \(x_k \in \mathcal{X}\) that has been reported most frequently by the oracles\footnote{In the case of a tie between two or more classes, we break the tie by selecting one of the classes uniformly at random. Other rules are applicable.} in $\mathcal{O}$.

Since $X$ belongs to a finite set, $\mathcal{X}$, the conditional PMF $p\left(y_n\big| x\right)$ for all $O_n^{(i)} \in \mathcal{O}_n^{\left(c_n\right)}$ can be represented by a $K \times K$ matrix known as the confusion matrix. The $(l,k)$-th element of this matrix is denoted by $p_{l|k}^{(n)}$ and is given by
\begin{equation}\label{eq:1116}
    p_{l|k}^{\left(n\right)}= \textrm{Pr}\left(Y_n=x_l|X=x_k \right).
\end{equation}

We assume that users have i.i.d. readings of $X$. As a result, $p_{l|k}^{\left(n\right)}$ does not depend on \(n\). Accordingly, we simplify our notation and denote the conditional probabilities in \eqref{eq:1116} as $p_{l|k}$ for all oracles in \(\mathcal{O}\). This scenario models users with i.i.d. observations of $X$. We leave the analysis of scenarios involving i.non-i.d. users for future work.

Next, we use the variable $p_{l|k}$ to introduce a property of the oracles that will be referenced later in the paper.

\begin{definition} \label{def:1}
Oracles are \textit{weakly accurate} if they are more likely to provide the correct value than an incorrect one, i.e.,
\begin{equation}
    p_{k|k}>p_{l|k}\,, \,\,\, \forall\, l,k \in [K],\text{~where~} l\neq k.
\end{equation}
\end{definition}

\subsection{Incentive Mechanism}
The IM distributes the total reward, $R$, among oracles that have reported an accurate value for the off-chain data point, $X$. In the absence of ground truth, the final output of the data-feed system, $\hat{X}$, is considered as the true value of $X$. More specifically, in the case of categorical data points, which is the main focus of this paper, only those oracles that have reported $\hat{X}$ receive a reward\footnote{When the off-chain data point is non-categorical, the reward-sharing mechanism may apply more complex criteria, such as the norm function, to evaluate the proximity of an oracle's reported value to $\hat{X}$, and allocate the corresponding reward proportionally to the accuracy of its observation.}.

To participate in the data-feeding process, each oracle is required to stake a minimum quantity of cryptocurrency, $s_{\min}$, as protection against Sybil attacks and to ensure compliance with the SLA. In the event of non-compliance with the obligations outlined in the SLA, the oracles may incur a loss of all or a portion of their stake. Without loss of generality, we assume $s_{\min}=1$ in the rest of this paper.

Like other permissionless decentralized systems, a more resourceful user, i.e., a user with greater staking power, will possess a higher capacity to influence the outcome of the decentralized data-feed system. We denote $s_{\max}$ as the threshold at which a user, controlling staking power surpassing this threshold, can independently determine the outcome of the AF, regardless of the values reported by other oracles. The exact value of $s_{\max}$ depends on both the relationship between the stake and the voting power of each oracle, as well as the accuracy of the oracles. For instance, in a decentralized data-feed system where oracles are highly accurate and for each oracle, the voting power is proportional to the stake, $s_{\max}$ approaches $50\%$ of the total staking power of the network. Since there is no known feasible solution to ensure the integrity and reliability of a decentralized data-feed system in the presence of a user controlling a staking power exceeding $s_{\max}$, we assume that all users hold less than $s_{\max}$ cryptocurrencies.

Let $s_n$ represent the total quantity of cryptocurrency that user $u_n$ possesses and can stake on its oracles. The quantity of cryptocurrency staked on oracle $O_n^{(i)}$ is represented by $s_n^{(i)}$, where
\begin{equation}
    \sum_{i=1}^{c_n} s_n^{(i)} \leq s_n.
\end{equation}

In the following, we adopt the model used in \cite{ChainLink-whitepaper} to formulate the IM employed in decentralized data-feed systems. 

Following each data-feeding task, the IM assigns \textit{reward factors} to each oracle in $\mathcal{O}$ based on the value of their stake and the value they submitted to the AF. Let \(r_n^{(i)}\) denote the reward factor of $O_n^{(i)}$. The actual reward (i.e., the payoff) of \(O_n^{(i)}\), denoted as \(R_n^{(i)}\left(\mathcal{O}\right)\), is then calculated by multiplying \(r_n^{(i)}\) by the total reward allocated for the task and normalizing the result by the sum of the reward factors of all participating oracles. Mathematically, this can be expressed as 
\begin{equation}\label{eq:111112}
R_n^{(i)} \left(\mathcal{O}\right)=\frac{r_n^{(i)}}{\sum_{m=1}^N \sum_{j=1}^{c_m}  r_m^{(j)}} R,
\end{equation}  
where the value of \(r_n^{(i)}\) is determined by a deterministic function as
\begin{equation}
   r_n^{(i)}= g\left(s_n^{(i)},Y_n^{(i)}, \hat{X} \right),
\end{equation} 
in which $s_n^{(i)}$, $Y_n^{(i)}$, and $\hat{X}$ represent, respectively, the quantity of cryptocurrency staked on $O_n^{(i)}$, the objective opinion of $O_n^{(i)}$ about $X$, and the final decision of the AF regarding the value of $X$. The function $g\left(s_n^{(i)}, Y_n^{(i)}, \hat{X}\right)$, which is modeled later, is the primary component through which the IM is constructed.

The payoff to user \(u_n\) is denoted by \(R_n\left(\mathcal{O}\right)\) and is given by
\begin{equation}
    R_n\left(\mathcal{O}\right) = \sum_{i=1}^{c_n} R_n^{(i)} \left(\mathcal{O}\right).
\end{equation}
Recalling that $\mathcal{O}=\mathcal{O}_n^{(c_n)} \cup \mathcal{O}_{- n}$, in the rest of this paper, we use the notation \(R_n\left(\mathcal{O}_n^{(c_n)}, \mathcal{O}_{- n} \right)\) to represent \(R_n\left(\mathcal{O}\right)\), highlighting the number of oracles controlled by $u_n$. 

At first glance, it may appear intuitive to assign a reward factor of one to an oracle if its observation of the off-chain data point agrees with \(\hat{X}\), and zero, otherwise. This is to discourage oracles from reporting inaccurate data. However, this methodology can incentivize rational users to create multiple oracles in an attempt to increase their payoff. As the marginal cost of creating an additional oracle is negligible, the profit increases at least linearly with the number of oracles each user creates. Therefore, this approach to determining the reward factors is ineffective and could compromise the accuracy and integrity of the entire data-feed system. To address this problem, existing decentralized data-feed systems distribute the total reward, \(R\), among the oracles that have reported \(\hat{X}\), proportionally based on the value of each oracle's stake \cite{ChainLink-whitepaper}, i.e., $r_n^{(i)}$ is given by
\begin{equation}\label{eq:1111}
    r_n^{(i)}= \begin{cases}
            s_n^{(i)} & \text{if~} Y_n^{(i)}= \hat{X} , \\
            0   & \textrm{otherwise}.
    \end{cases}
\end{equation}
However, as we show in the following section, using \eqref{eq:1111} as the IM, oracles remain strongly incentivized to distribute their staking power across multiple oracles that report the same value, in order to maximize their payoffs. Hence, $r_n^{(i)}$ defined in \eqref{eq:1111} is not an IM that can address the mirroring attack.

\section{Proposed Incentive Mechanism for Mitigating the Mirroring Attack}
To propose the IM, i.e., to propose $g\left(s_n^{(i)},Y_n^{(i)}, \hat{X} \right)$, we use insights from the results in \cite{ourpaper}. The authors in \cite{ourpaper} derived an upper bound on the error rate of the MVF for the general multi-class classification problem. They showed that if individual voters satisfy the following conditions, then the error rate of the MVF exponentially decays toward zero:
\begin{itemize}
    \item \textbf{Condition 1:} Oracles are weakly accurate, as defined in Definition \ref{def:1}.
    \item \textbf{Condition 2:} Oracles are mutually independent, which means that given the true classification of data points $X$, the conditional PMFs of the output of any two oracles are independent of each other.
\end{itemize}

When Condition 1 is not met, the error rate of the MVF increases as the number of mutually independent oracles increases \cite{ourpaper}, making it impossible to build an accurate decentralized data-feed system. On the other hand, when Condition 1 is satisfied, ensuring that the oracles are independent becomes sufficient for building an accurate decentralized data-feed system. Therefore, we only focus on scenarios where Condition 1 holds and aim to address the challenge of meeting Condition 2.

While satisfying Condition 1 is often feasible in practical applications, meeting Condition 2 remains a challenge. This is because the existing IMs encourage users to implement the mirroring attack by replicating their oracles, thereby violating Condition 2.

Next, we formally define the mirroring attack and present our solution to mitigate it, ensuring that users are effectively incentivized to satisfy Condition 2.

\subsection{Mirroring Attack}
We start with the definition of the mirroring attack.
\begin{definition}
The mirroring attack is a type of Sybil attack in which a user intentionally joins the data-feed system with multiple oracle nodes and submits identical values for a given off-chain data query through each of these oracles.
\end{definition}

The mirroring attack can introduce errors into the data-feed system by undermining the intended mutual independence of the oracles. In open and decentralized data-feed systems that allow anonymous participation of oracles, it is nearly impossible to detect and prevent mirroring attacks. This is because there is no reliable methodology for the AF to measure the correlations between oracles and their respective controlling users.

By employing the mirroring attack, a user can amplify their influence on the final output of the AF, thereby increasing the probability of receiving rewards for each off-chain data query task. For example, consider a binary data query task where $\lfloor\frac{N-1}{2}\rfloor$ oracles, including $O_n^{(1)}$, voted for $x_1$, while $\lceil\frac{N+1}{2}\rceil$ oracles voted for $x_2$. In this scenario, the MVF would output $x_2$, resulting in no reward for $u_n$, the user controlling $O_n^{(1)}$. However, if the owner of $O_n^{(1)}$ had implemented the mirroring attack by replicating its oracle node and engaging in the protocol with three oracles instead of one, the MVF would output $x_1$, resulting in a reward for $u_n$. Therefore, even with the reward factor given in \eqref{eq:1111}, rational users are incentivized to use the mirroring attack to increase their expected reward.

\subsection{Superlinear Incentive Mechanism}

We propose an alternative IM to the one in \eqref{eq:1111}, which we call the \textit{superlinear IM}, to discourage the mirroring attack. The proposed IM calculates the reward factor of each oracle using a function that grows faster than linear with respect to the oracle’s stake. This approach incentivizes rational users to consolidate all their staking power on a single oracle to maximize their reward factor. More specifically, we use a power function due to its simplicity and its ability to control the degree of superlinearity through a single parameter, making it a practical choice for implementation and analysis. The reward factor for the proposed IM is expressed as 
\begin{equation}\label{eq:1112}
    r_{n}^{(i)}= \begin{cases}
            \left(s_{n}^{(i)}\right)^d & \text{if~} Y_n^{(i)}= \hat{X}, \\
            0   & \textrm{otherwise},
    \end{cases}
\end{equation}
where $d>1$ determines the rate at which the reward factor increases with the oracle's stake. The parameter $d$ in \eqref{eq:1112} must be sufficiently large to ensure that, for all users, the expected reward is maximized by staking all their cryptocurrency on a single oracle. However, an excessively high value of $d$ could lead to an exponential increase in the payoff for wealthy users, potentially compromising incentive fairness and, as a result, reducing the system's level of decentralization. Specifically, as we increase $d$ (i.e., $d\to \infty$), the mirroring attack becomes less attractive; however, at the same time, the reward distribution becomes highly centralized, with a few wealthy users capturing almost the entire reward. Therefore, our objective is to design and fine-tune the system parameter $d$ and evaluate the effectiveness of the proposed mechanism in mitigating the mirroring attack while preventing wealthy users from compromising decentralization. More specifically, we aim to demonstrate the ability of the IM in \eqref{eq:1112} to establish the desired behavior where participating in the protocol with a single oracle becomes a Nash Equilibrium strategy for all users. This is done in the following.

\subsection{Game Theoretic Analysis of the Proposed Incentive Mechanism}\label{game-theoretic analysis}
Let $d_{\text{opt}}$ denote the minimum value of $d$ in \eqref{eq:1112} that is sufficient to incentivize all rational users to participate in the protocol with exactly one oracle. In other words, $d_{\text{opt}}$ represents the smallest value of \(d\) that makes the strategy profile \(\mathcal{O}^*= \left\{O_1^*, O_2^*, \dots, O_n^* \right\}\), where each user stakes all their cryptocurrency on a single oracle, a Nash Equilibrium. Expressing this mathematically, \(d= d_{\text{opt}}\) is the minimum value of $d$ for which we have
\begin{multline}\label{eq:1118}
    E\left\{R_n\left(\mathcal{O}_n^{(c_n)} , \mathcal{O}_{- n}^* \right) \right\}\leq E\left\{R_n\left(\mathcal{O}_n^{*} , \mathcal{O}_{- n}^* \right)\right\}, \\ \forall \, n \in [N] \textrm{ and } \forall \, c_n \in [s_n].
\end{multline}

\sloppy Our focus in the rest of this section is on deriving \(E\left\{R_n\left(\mathcal{O}_n^{(c_n)}, \mathcal{O}_{-n}^* \right)\right\}\) as a function of $d$. We will then utilize this derivation to compute $d_{\textrm{opt}}$ in the next section.

\begin{theorem} \label{theorem:1.5}
    For any user $u_n$ employing the mirroring attack by using $c_n$ oracles, we have
    \begin{flalign}\label{eq:11111}
        &\max_{\mathbf{s_n^{\left(1\right)}, s_n^{\left(2\right)}, \dots, s_n^{\left(c_n\right)}}} E\left\{R_n\left(\mathcal{O}_n^{(c_n)} , \mathcal{O}_{- n} \right) \right\} \nonumber\\
        &\textrm{~~~~~~~}=  \sum_{\mathbf{y} \in \left\{\mathbf{y}: \mathbf{y} \in \mathcal{X}^{M}, f\left(\mathbf{y}\right) = Y_n \right\}} \textrm{Pr}\left(\mathbf{Y}=\mathbf{y} \right)&&\nonumber\\ 
        &\textrm{~~~~~~~~~~~~~~~~~~~~}\times\frac{c_n-1+ \left(s_n- c_n+1\right) ^d}{\left(c_n-1 + \left(s_n- c_n +1\right) ^d \right)+  \sum_{\substack{m=1 \\ m\neq n}}^N {r_m}} ,&&
    \end{flalign}
 where $r_m=\sum_{i=1}^{c_m}r_m^{\left(i\right)}$ and denotes the aggregated reward factor of oracles owned by user $u_m$. Moreover, $M = \sum_{n=1}^N c_n$ represents the total number of oracles, $Y_n$ denotes the value submitted by the oracles owned by $u_n$, i.e., $Y_n= Y_n^{(1)}= Y_n^{(2)}= \dots = Y_n^{(c_n)}$, and $f\left(\mathbf{y}\right) = Y_n$ denotes that the output of the MVF is $\hat{X}= Y_n$ given that~$\mathbf{Y}=\mathbf{y}$.   
\end{theorem}
\begin{IEEEproof}
    Utilizing \eqref{eq:111112} with the reward factor defined in \eqref{eq:1112}, we have
\begin{align}\label{eq:2112}
    &E\left\{R_n\left(\mathcal{O}_n^{(c_n)} , \mathcal{O}_{- n} \right) \right\} \nonumber\\
    &\textrm{~~~~~~~~~~~~~~~}=E\left\{\frac{r_n}{\sum_{m=1}^N {r_m}} \right\} R_t \nonumber\\ 
    &\textrm{~~~~~~~~~~~~~~~}\stackrel{(a)}= \sum_{\mathbf{y} \in \left\{\mathbf{y}: f\left(\mathbf{y}\right) = Y_n \right\}} \textrm{Pr}\left(\mathbf{Y}=\mathbf{y} \right) \frac{r_n}{\sum_{m=1}^N {r_m}} \nonumber\\
    &\textrm{~~~~~~~~~~~~~~~}\textrm{~~~~~}+ \sum_{\mathbf{y} \in \left\{\mathbf{y}: f\left(\mathbf{y}\right) \neq Y_n \right\}} \textrm{Pr}\left(\mathbf{Y}=\mathbf{y} \right) \frac{r_n}{\sum_{m=1}^N {r_m}}  \nonumber\\
    &\textrm{~~~~~~~~~~~~~~~}\stackrel{(b)}=  \sum_{\mathbf{y} \in \left\{\mathbf{y}: f\left(\mathbf{y}\right) = Y_n \right\}} \textrm{Pr}\left(\mathbf{Y}=\mathbf{y} \right) \frac{r_n}{\sum_{m=1}^N {r_m}}\nonumber\\
    &\textrm{~~~~~~~~~~~~~~~}\stackrel{(c)}=  \sum_{\mathbf{y} \in \left\{\mathbf{y}: f\left(\mathbf{y}\right) = Y_n \right\}} \textrm{Pr}\left(\mathbf{Y}=\mathbf{y} \right) \nonumber\\ 
    &\textrm{~~~~~~~~~~~~~~~}\textrm{~~~~~~~~~~~~~~~~~~~~~}\times \frac{\sum_{i=1}^{c_n} \left(s_{n}^{(i)}\right)^d}{\sum_{i=1}^{c_n} \left(s_{n}^{(i)}\right)^d +  \sum_{\substack{m=1 \\ m\neq n}}^N {r_m}},
\end{align}
    where $(a)$ is derived from the law of total expectation and the assumption that $R_t=1$. Moreover, $(b)$ is due to the fact that if $Y_n\neq f \left(\mathbf{Y} \right)$, then $r_n =0$. Finally, $(c)$ is obtained by substituting $r_n$ using~\eqref{eq:1112}.
    
    Next, we show that \eqref{eq:2112} can be further simplified to obtain \eqref{eq:11111}. To this end, let $\mathcal{S}_n^*$ denote the set of all staking strategies that $u_n$ can employ to obtain the expected payoff in the right-hand side of \eqref{eq:11111}. We first show that $\mathcal{S}_n^* \neq \emptyset$. This is because by staking $s_n-\left(c_n-1\right)$ units of cryptocurrency on one of its oracles while staking exactly $s_{\min}=1$ cryptocurrency on each one of the remaining $c_n-1$ oracles, $u_n$ can attain the expected payoff in~\eqref{eq:11111}. Next, we employ a proof by contradiction to demonstrate that $E\left\{R_n\left(\mathcal{O}_n^{(c_n)}, \mathcal{O}_{- n} \right) \right\}$ cannot take values larger than the right-hand side of \eqref{eq:11111}. This is done in the following.
    
    Let $\mathbf{s_n}^{'} =\left(s_n^{(1)^{'}}, s_n^{(2)^{'}}, \dots, s_n^{(c_n)^{'}} \right)$ be the optimal strategy that maximizes $E\left\{R_n\left(\mathcal{O}_n^{(c_n)}, \mathcal{O}_{- n} \right) \right\}$ to a value larger than the right-hand side of \eqref{eq:11111}. There are at least two elements in $\mathbf{s_n}^{'}$ that are greater than $s_{\min}=1$, as otherwise, $\mathbf{s_n}^{'} \in \mathcal{S}_n^*$ and the expected reward would be equal to the right-hand side of \eqref{eq:11111}. Without loss of generality, let $s_n^{(1)^{'}}$ and $s_n^{(2)^{'}}$ be two such elements, where $1<s_n^{(1)^{'}}\leq s_n^{(2)^{'}}$. Due to the convexity of the reward factor in \eqref{eq:1112}, we can always increase $E\left\{R_n\left(\mathcal{O}_n^{(c_n)}, \mathcal{O}_{- n} \right) \right\}$ by reducing $\epsilon$ units from $s_n^{(1)}$ and adding it to $s_n^{(2)}$, where $0< \epsilon \leq s_n^{(1)^{'}}-1$. Therefore, $\mathbf{s_n}^{'}$ cannot be the optimal strategy, and the proof is complete.
\end{IEEEproof}
Theorem \ref{theorem:1.5} derives the maximum value of $E\left\{R_n\left(\mathcal{O}_n^{(c_n)}, \mathcal{O}_{- n} \right) \right\}$ over all feasible strategies that $u_n$ can employ to stake on its oracles.

\begin{corollary}\label{rem:2}
Since users are assumed to be rational, they always adopt the staking strategy that leads to maximizing their expected payoff. Therefore, the average reward for each user is given by \eqref{eq:11111}, i.e., 
\begin{align}\label{eq:111113}
    &E\left\{R_n\left(\mathcal{O}_n^{(c_n)} , \mathcal{O}_{- n} \right) \right\} \nonumber\\
    &\textrm{~}= \max_{\mathbf{s_n^{\left(1\right)}, s_n^{\left(2\right)}, \dots, s_n^{\left(c_n\right)}}} E\left\{R_n\left(\mathcal{O}_n^{(c_n)} , \mathcal{O}_{- n} \right) \right\}  \nonumber\\
    &\textrm{~}= \sum_{\mathbf{y} \in \left\{\mathbf{y}: f\left(\mathbf{Y}\right) = Y_n \right\}} \textrm{Pr}\left(\mathbf{Y}=\mathbf{y} \right) \nonumber\\
    & \textrm{~~~~}\times\frac{c_n-1+ \left(s_n- c_n+1\right) ^d}{\left(c_n-1 + \left(s_n- c_n +1\right) ^d \right)+  \sum_{\substack{m=1 \\ m\neq n}}^N {r_m}} \,
    , \forall\, n \in [N].
\end{align}    

\end{corollary}

In the next corollary, we derive the expected payoff of the user $u_n$ in the special case where all other users follow the desired strategy of concentrating their entire stake on a single oracle. Later in this section, we use this result to determine the optimal value of the parameter $d$.

\begin{corollary} \label{theorem:2}
    The expected payoff for user \(u_n\) controlling \(c_n\) oracles with total staking power $s_n$, while all other users concentrate all their staking power on a single oracle, is given by
    \begin{align}\label{eq:1114}
        &E\left\{R_n\left(\mathcal{O}_n^{(c_n)}, \mathcal{O}_{-n}^* \right)\right\} \nonumber\\
        &\textrm{~~}= \sum_{\mathbf{y} \in \left\{\mathbf{y}: f\left(\mathbf{y}\right) = Y_n \right\}} \textrm{Pr}\left(\mathbf{Y}=\mathbf{y} \right) \nonumber\\
        &\textrm{~~~~}\times\frac{c_n-1+ \left(s_n- c_n+1\right) ^d}{\left(c_n-1 + \left(s_n- c_n +1\right)^d \right)+  \sum_{m \in \mathcal{Y}_{correct\setminus n}} {s_m}^d},
    \end{align}
where $\mathcal{Y}_{correct\setminus n} = \left\{m: m\in [N] \setminus n, Y_m=f\left(\mathbf{y}\right) \right\}$. This is obtained by substituting $r_m$-s in \eqref{eq:111113} with the definition provided in \eqref{eq:1112} when $c_m=1, \, \forall \, m\in [N]\setminus n$.  
\end{corollary}

In the next section, we utilize \eqref{eq:1114} to compute $d_{\textrm{opt}}$, the minimum value of $d$ that satisfies \eqref{eq:1118}.

\subsection{Obtaining $d_{\text{opt}}$ for the Proposed Incentive Mechanism}
In the following, we introduce a brute-force algorithm to find the minimum value of the parameter $d$ that maximizes the expected reward for every user $u_n$ when users consolidate all their stake power on a single oracle, i.e., 
\begin{equation} \label{eq:1113}
    \underset{c_n}{\textrm{argmax}} \left(E\left\{R_n\left(\mathcal{O}_n^{(c_n)}, \mathcal{O}_{-n}^* \right)\right\}\right) =1, \,\,\,\,\,\,\, \forall \, n \in [N].
\end{equation}
The proposed algorithm takes as input the distribution of the staking power among independent users, $\mathbf{S}=\left( s_1, s_2, \dots, s_N \right)$, along with $\mathbf{p}$, the confusion matrix defined in \eqref{eq:1116}. However, in practice, the data-feed smart contract has no means of identifying actual users who may control one or more oracles. As a result, the staking power of users, $\mathbf{S}$, remains unknown to the data-feed smart contract, making the proposed algorithm impractical. Later, we modify this algorithm to enable determining $d_{\textrm{opt}}$ without knowing the staking power of the users.

\begin{algorithm}
\SetKwInOut{Input}{input}\SetKwInOut{Output}{output}
\caption{Find Optimal Value of $d$}
\label{alg:1}
\Input{$\epsilon$, $s_1$, $s_2$, $\dots$, $s_N$, $\mathbf{p}$}
\Output{$d_{\text{opt}}$}
\BlankLine
 $d \gets 1$\;
\While{Stop-Loop = \textrm{False}}
{
    \For{$n \gets 1$ to $N$}
        {
         Compute $E\left\{R_{n} \left(\mathcal{O}_n^*, \mathcal{O}_{-n}^*\right)\right\}$ using \eqref{eq:1114}\;
        }
    \For{$n \gets 1$ to $N$}
    {
        \For{$c_n \gets 2$ to $s_n$}
        {
             Compute $E\left\{R_{n}\left(\mathcal{O}_n^{(c_n)}, \mathcal{O}_{-n}^*\right)\right\}$ using \eqref{eq:1114}\;
            \If{$E\left\{R_{n}\left(\mathcal{O}_n^*, \mathcal{O}_{-n}^*\right)\right\} < E\left\{R_{n}\left(\mathcal{O}_n^{(c_n)}, \mathcal{O}_{-n}^*\right)\right\}$}
            {
                 $d \gets d + \epsilon$\;
            }

        }
    }
    $d_{\text{opt}} \gets d$\;
    $Stop-Loop = \textrm{True}$\;
}
\end{algorithm}

Starting from $d=1$, Algorithm \ref{alg:1} increments $d$ until the inequality in \eqref{eq:1118} is satisfied for all $n \in [N]$ and for all $c_n \in [s_n]$. The algorithm then outputs the resulting value as $d_{\textrm{opt}}$. It is important to note that in this algorithm, $c_n$ is a variable that can range from $2$ to $s_n$.

Algorithm \ref{alg:1} finds $d_{\text{opt}}$ in polynomial time. This is because calculating $E\left\{R_{n}\left(\mathcal{O}_n^*, \mathcal{O}_{-n}^*\right)\right\}$ using \eqref{eq:1114} requires computing the full Multinomial distribution of oracle submissions over the set $\mathcal{X}=\left\{x_1,x_2, \dots, x_K\right\}$, which has a time complexity of $O\left( {M+K-1 \choose K-1} \right)= O \left(M^{\left(K-1\right)}\right)$. Therefore, the time complexity of computing $E\left\{R_{n}\left(\mathcal{O}_n^*, \mathcal{O}_{-n}^*\right)\right\}$ for all $n \in N$ and all $c_n \in [s_n]$ is $O\left(\left(\sum_{n=1}^N s_n\right) M^{\left(K-1\right)}\right)$.

In practice, the data-feed smart contract cannot identify individual users and their respective staking power, making the implementation of Algorithm \ref{alg:1} challenging. However, the following theorem demonstrates that we can use the stakes of oracles as inputs to Algorithm \ref{alg:1}, instead of the staking power of users, and still obtain the same output for $d_{\textrm{opt}}$. This is important because, while the data-feed system may not have visibility into the staking power of individual users, it can always observe the quantity of cryptocurrency staked on each oracle.

\begin{theorem} \label{theorem:3}
    Let $\mathbf{S'}=\left(s'_1, s'_2, \dots, s'_M\right)$ denote the vector representing the quantities of cryptocurrency staked on oracles, where $M = \sum_{n=1}^N c_n$ is the total number of oracles in the network and $s'_m$ denotes the quantity of cryptocurrency staked on the $m$-th oracle, with oracles being randomly labeled. Substituting the vector $\mathbf{S}=\left(s_1, s_2, \dots, s_N\right)$ with $\mathbf{S'}=\left(s'_1, s'_2, \dots, s'_M\right)$ in Algorithm 1 does not change the output of the algorithm.
\end{theorem}
\begin{IEEEproof}
    We first consider the case where users follow the favored behavior and participate in the protocol with a single oracle node each. In this scenario, we have $\mathbf{S'}=\mathbf{S}$ and the output of Algorithm \ref{alg:1} evidently satisfies \eqref{eq:1113}. Therefore, the proof is complete for this case. Next, we show that this is the only possible scenario. 
    
    We prove through contradiction that when users are rational and Algorithm \ref{alg:1} is applied to determine $d_{\textrm{opt}}$, $\mathbf{S'}$ cannot differ from $\mathbf{S}$. Let us assume that for a given $d_{\textrm{opt}}$ obtained through Algorithm \ref{alg:1}, user $u_n$ with staking power $s_n$ is incentivized to implement the mirroring attack. Considering that all users are rational, for this attack to be feasible, $s_n$ must exceed $s_0=\max\left(s'_1, s'_2, \dots, s'_M \right)$. This is because Algorithm \ref{alg:1} determines $d_{\textrm{opt}}$ so that all users with staking power in $[s_0]$ are encouraged to concentrate their staking power on a single oracle.
    However, even when $s_n> s_0$, $u_n$ can increase its expected reward by concentrating its staking power on a single oracle. This occurs due to the design of Algorithm \ref{alg:1}, where staking $s_n$ on a single oracle leads to increasing $s_0$ and, consequently, $d_{\textrm{opt}}$. As a result, and due to the convexity of the reward factor in \eqref{eq:1112}, the increase in $d_{\textrm{opt}}$ increases the reward factor of $u_n$ faster than all other users. Therefore, $u_n$ can't be incentivized to implement the mirroring attack. Hence, $\mathbf{S}= \mathbf{S'}$ and the proof is complete.
\end{IEEEproof}

Theorem \ref{theorem:3} shows that the IM can utilize the stake quantities of oracles as inputs to Algorithm \ref{alg:1} and obtain $d_{\textrm{opt}}$. Another parameter that the data-feed smart contract needs for running Algorithm~\ref{alg:1} is the confusion matrix of the oracles. Given that the oracles are presumed to have weak accuracy and that their outputs are identically distributed given the true value of the off-chain data point, we can empirically estimate the confusion matrix $\mathbf{P}$ using the historical values submitted by the oracles, while taking the final outputs of the data-feed system as the ground truth.

\section{Numerical Results}
To show the performance of the proposed IM, we consider a decentralized data-feed system consisting of $10$ users, $u_1,u_2,\dots,u_{10}$, where their staking power is given by 
\begin{align*}
    s_1=8, \, s_2=5,\, s_3=3,\, s_4=8,\, s_5=4,\, \\ s_6=7,\, s_7=6,\, s_8=5,\, s_9=7,\, s_{10}=2.
\end{align*}
Furthermore, we assume that the minimum stake required for participation in the protocol is $s_{\min}=1$.

We use the following confusion matrix in our simulations
\begin{equation*}
\mathbf{P}=\begin{bmatrix}
0.6439 & 0.0705 & 0.0354 & 0.0592 & 0.1910 \\
0.1009 & 0.5070 & 0.0451 & 0.1103 & 0.2367 \\
0.0611 & 0.0930 & 0.5749 & 0.0783 & 0.1927 \\
0.0333 & 0.0549 & 0.0475 & 0.8072 & 0.0571 \\
0.1147 & 0.2231 & 0.0550 & 0.1598 & 0.4474 \\
\end{bmatrix}.
\end{equation*}
This matrix was derived from a real-world dataset provided by Amazon Mechanical Turk~\cite{dataSet}. The dataset consists of 6000 sentiment classifications of 300 tweets, with gold-standard sentiment labels provided by 110 human annotators~\cite{dataSet}. To ensure data quality, we removed parts of the dataset that did not include a gold-standard label and data provided by oracles who participated in less than $10\%$ of the tasks. We then calculated the average accuracy of the oracles by comparing their opinions with the gold-standard label, which allowed us to obtain the confusion matrix.

In our simulations, we investigate the impact of $u_1$ employing the mirroring attack by participating with $c_1$ oracles. We assume the best-case scenario where the other users comply with the desired behavior and participate with only one oracle each.

\begin{figure} [htbp]
\centering
\includegraphics[width=0.52\textwidth]{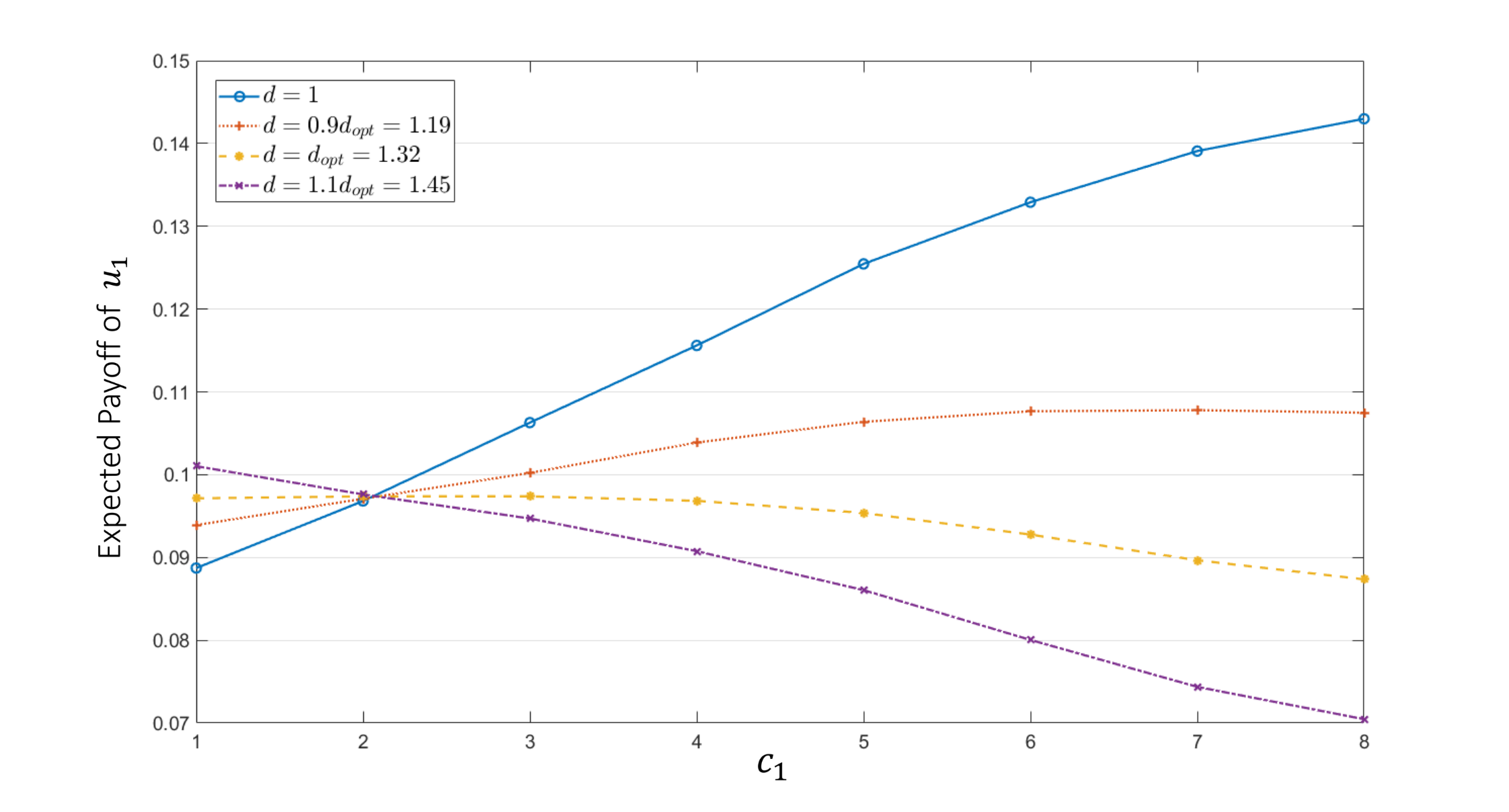}
\caption{The expected payoff to $u_1$ as a proportion of $R_t$ for different values of $c_1$ and $d$. 
}\label{fig:reward}
\end{figure}

Figure \ref{fig:reward} illustrates the expected payoff to $u_1$ versus the number of oracles controlled by $u_1$ for various values of the parameter $d$. 
As illustrated in Figure \ref{fig:reward}, when $d=1$, user $u_1$ is incentivized to participate with as many oracles as its staking power allows ($c_1=8$). This scenario represents a special case where the proposed IM in \eqref{eq:1111} reduces to the IM in \eqref{eq:1112}, i.e., the existing IM implemented in the data-feed systems. This observation shows the ineffectiveness of the existing IMs in addressing the mirroring attack.

As shown in Figure \ref{fig:reward}, in the current IMs of decentralized data-feed systems, when $1<d<d_{\textrm{opt}}$, $u_1$ is still motivated to participate with more than one oracle, however, the gain it can achieve through implementing the mirroring attack decreases compared to the scenario where $d=1$. When $d \geq d_{\textrm{opt}}$, the IM effectively encourages rational users to concentrate their entire staking power on a single oracle. However, for $d>d_{\textrm{opt}}$, the fairness of the IM decreases as users with greater staking power attract exponentially higher payoffs, which in turn leads to an increase in the centralization of the system over time. The critical threshold where $d=d_{\textrm{opt}}$, represents the point where the expected payoff to $u_1$, the user with the highest staking power in our example, is minimized and approaches as close as possible to its proportion of the total staking power across all users, ensuring the best feasible incentive fairness and system decentralization.

\begin{figure} [htbp]
\centering
\includegraphics[width=0.52\textwidth]{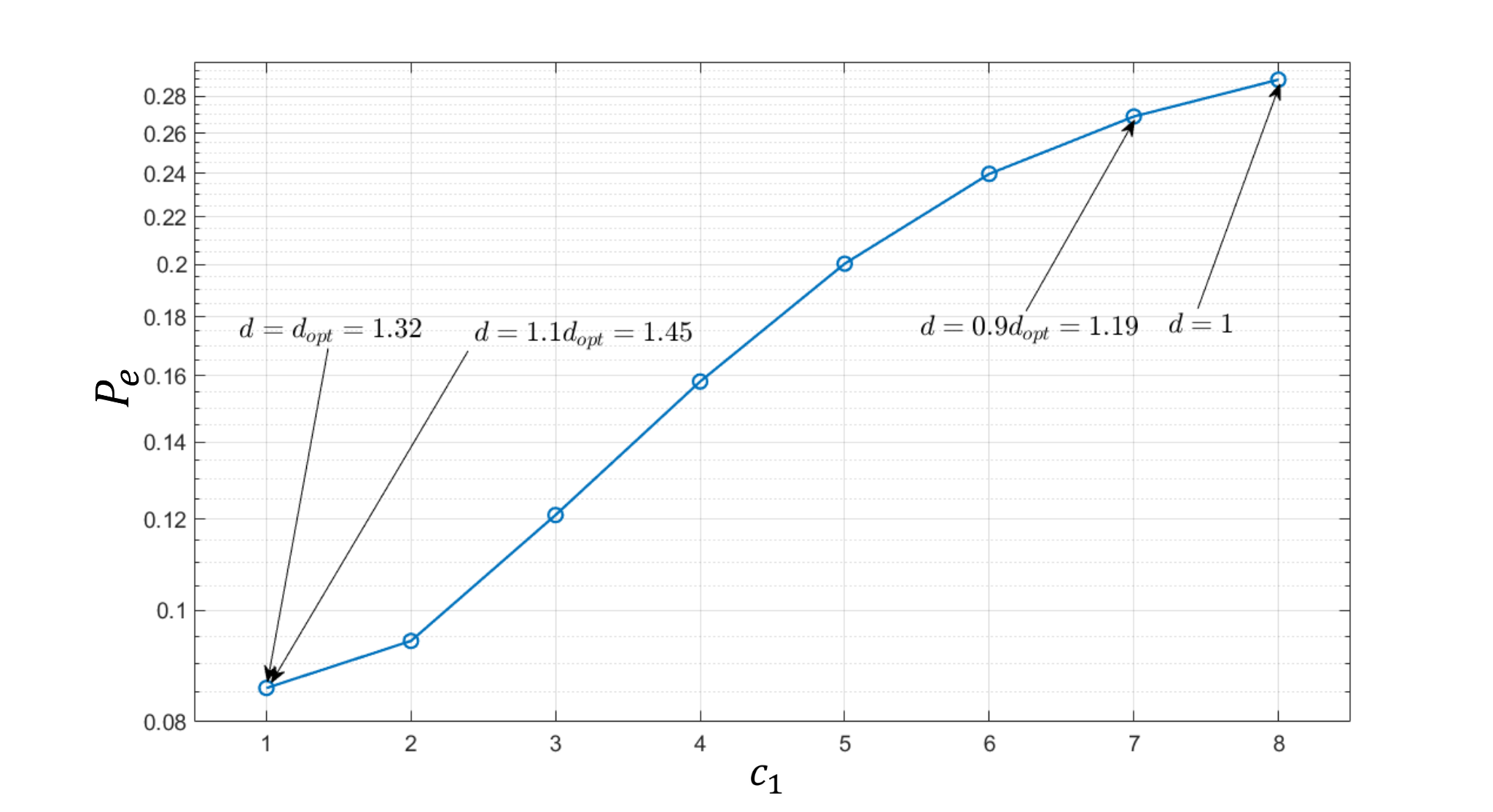}
\caption{The error rate of the considered decentralized data-feed system for different values of $c_1$ and $d$. 
}\label{fig:error}
\end{figure}

Figure \ref{fig:error} illustrates the accuracy of the data-feed system as a function of the number of oracles controlled by $u_1$. 

As shown in Figure \ref{fig:error}, the error rate of the decentralized data-feed system increases as the number of oracles controlled by $u_1$ increases. This is because the mutual dependency of the values reported by oracles increases with $c_1$, leading to a reduction in the accuracy of the MVF \cite{ourpaper}. This scenario reflects the outcome when $d=1$. 

On the other hand, the scenario where $u_1$ participates with only one oracle ($c_1=1$) yields the lowest error rate. As we previously discussed, this is the scenario that reflects $d=d_{\textrm{opt}}$. More specifically, when $d=d_{\textrm{opt}}$, the strategy of concentrating the entire staking power on a single oracle forms a Nash Equilibrium for all users. It is important to note that increasing the value of $d$ beyond $d_{\textrm{opt}}$, does not lead to any further reduction in the error rate.

\section{Conclusion}
We presented a new IM to mitigate the mirroring attack in a decentralized data-feed system that employs the MVF as an AF. Through a game-theoretic analysis, we demonstrated the effectiveness of the proposed IM in discouraging users from employing the mirroring attack. More specifically, we showed that when initiated with a properly chosen system parameter, the proposed IM establishes participation with a single oracle node as a Nash Equilibrium strategy for all users.

In essence, the proposed IM exponentially increases the reward factor of each oracle as their stake grows in order to encourage users to consolidate all their staking power on a single oracle. The slope by which the reward factor grows with increasing stake is controlled by a system parameter denoted as $d$. We introduced an algorithm that utilizes the vector of stakes of all oracles and their historical submissions to determine the optimal value of $d$. This algorithm ensures that the value of $d$ is large enough to cancel out the effect of replicating the same value through multiple oracles, which could otherwise increase the probability of swaying the outcome of the MVF. Moreover, the algorithm ensures that the lowest possible value for $d$ is selected, as choosing an unnecessarily high value for $d$ can discourage users with low staking power from participating in the protocol.

We believe that the solution proposed in this paper for addressing the mirroring attack, coupled with the recent results on the accuracy of the MVF \cite{ourpaper}, plays a crucial role in ensuring the reliability and accuracy of decentralized data-feed systems. Future work could explore extensions of these results to independent and non-identically distributed oracles and data-feed systems for continuous off-chain data points. Moreover, a broader range of superlinear functions, beyond the power function, could be explored for determining the reward factor. Furthermore, future works may investigate solutions for reducing the time complexity of Algorithm \ref{alg:1} to enhance its applicability in networks with a substantial number of oracles.

\ifCLASSOPTIONcaptionsoff
  \newpage
\fi

\bibliographystyle{IEEEtran}
\bibliography{references}

\begin{thebibliography}{10}
\providecommand{\url}[1]{#1}
\csname url@samestyle\endcsname
\providecommand{\newblock}{\relax}
\providecommand{\bibinfo}[2]{#2}
\providecommand{\BIBentrySTDinterwordspacing}{\spaceskip=0pt\relax}
\providecommand{\BIBentryALTinterwordstretchfactor}{4}
\providecommand{\BIBentryALTinterwordspacing}{\spaceskip=\fontdimen2\font plus
\BIBentryALTinterwordstretchfactor\fontdimen3\font minus \fontdimen4\font\relax}
\providecommand{\BIBforeignlanguage}[2]{{%
\expandafter\ifx\csname l@#1\endcsname\relax
\typeout{** WARNING: IEEEtran.bst: No hyphenation pattern has been}%
\typeout{** loaded for the language `#1'. Using the pattern for}%
\typeout{** the default language instead.}%
\else
\language=\csname l@#1\endcsname
\fi
#2}}
\providecommand{\BIBdecl}{\relax}
\BIBdecl

\bibitem{8494045}
B.~K. Mohanta, S.~S. Panda, and D.~Jena, ``An overview of smart contract and use cases in blockchain technology,'' in \emph{2018 9th International Conference on Computing, Communication and Networking Technologies (ICCCNT)}, 2018, pp. 1--4.

\bibitem{khan2021blockchain}
S.~N. Khan, F.~Loukil, C.~Ghedira-Guegan, E.~Benkhelifa, and A.~Bani-Hani, ``Blockchain smart contracts: Applications, challenges, and future trends,'' \emph{Peer-to-peer Networking and Applications}, pp. 1--25, 2021.

\bibitem{Oracle-Review}
H.~{Al-Breiki}, M.~H.~U. {Rehman}, K.~{Salah}, and D.~{Svetinovic}, ``Trustworthy blockchain oracles: Review, comparison, and open research challenges,'' \emph{IEEE Access}, vol.~8, pp. 85\,675--85\,685, 2020.

\bibitem{town}
\BIBentryALTinterwordspacing
F.~Zhang, E.~Cecchetti, K.~Croman, A.~Juels, and E.~Shi, ``Town crier: An authenticated data feed for smart contracts,'' in \emph{Proceedings of the 2016 ACM SIGSAC Conference on Computer and Communications Security}, ser. CCS '16.\hskip 1em plus 0.5em minus 0.4em\relax New York, NY, USA: Association for Computing Machinery, 2016, p. 270–282. [Online]. Available: \url{https://doi.org/10.1145/2976749.2978326}
\BIBentrySTDinterwordspacing

\bibitem{Ritzdorf2017TLSNNO}
H.~Ritzdorf, K.~W{\"u}st, A.~Gervais, G.~Felley, and S.~Capkun, ``Tls-n: Non-repudiation over tls enabling - ubiquitous content signing for disintermediation,'' \emph{IACR Cryptol. ePrint Arch.}, vol. 2017, p. 578, 2017.

\bibitem{DECO}
\BIBentryALTinterwordspacing
F.~Zhang, D.~Maram, H.~Malvai, S.~Goldfeder, and A.~Juels, ``Deco: Liberating web data using decentralized oracles for tls,'' in \emph{Proceedings of the 2020 ACM SIGSAC Conference on Computer and Communications Security}, ser. CCS '20.\hskip 1em plus 0.5em minus 0.4em\relax New York, NY, USA: Association for Computing Machinery, 2020, p. 1919–1938. [Online]. Available: \url{https://doi.org/10.1145/3372297.3417239}
\BIBentrySTDinterwordspacing

\bibitem{ASTRAEA}
J.~{Adler}, R.~{Berryhill}, A.~{Veneris}, Z.~{Poulos}, N.~{Veira}, and A.~{Kastania}, ``Astraea: A decentralized blockchain oracle,'' in \emph{2018 IEEE International Conference on Internet of Things (iThings) and IEEE Green Computing and Communications (GreenCom) and IEEE Cyber, Physical and Social Computing (CPSCom) and IEEE Smart Data (SmartData)}, 2018, pp. 1145--1152.

\bibitem{ChainLink-whitepaper}
\BIBentryALTinterwordspacing
S.~Ellis, A.~Juels, and S.~Nazarov. (2017) Chainlink: A decentralized oracle network. [Online]. Available: \url{https://link.smartcontract.com/whitepaper}
\BIBentrySTDinterwordspacing

\bibitem{Witnet}
\BIBentryALTinterwordspacing
A.~S.~D. Pedro, D.~Levi, and L.~I. Cuende, ``\BIBforeignlanguage{en}{Witnet: A decentralized oracle network protocol},'' 2017. [Online]. Available: \url{http://rgdoi.net/10.13140/RG.2.2.28152.34560}
\BIBentrySTDinterwordspacing

\bibitem{SoKOracle}
\BIBentryALTinterwordspacing
S.~Eskandari, M.~Salehi, W.~C. Gu, and J.~Clark, \emph{SoK: Oracles from the Ground Truth to Market Manipulation}.\hskip 1em plus 0.5em minus 0.4em\relax New York, NY, USA: Association for Computing Machinery, 2021, p. 127–141. [Online]. Available: \url{https://doi.org/10.1145/3479722.3480994}
\BIBentrySTDinterwordspacing

\bibitem{Wang-TDSC-Strategic}
D.~Wang, T.~Muller, and J.~Sun, ``Provably secure decisions based on potentially malicious information,'' \emph{IEEE Transactions on Dependable and Secure Computing}, vol.~21, no.~5, pp. 4388--4403, 2024.

\bibitem{Rezvani-TDSC-Colllusion}
M.~Rezvani, A.~Ignjatovic, E.~Bertino, and S.~Jha, ``Secure data aggregation technique for wireless sensor networks in the presence of collusion attacks,'' \emph{IEEE Transactions on Dependable and Secure Computing}, vol.~12, no.~1, pp. 98--110, 2015.

\bibitem{576970}
L.~Lam and C.~Suen, ``A theoretical analysis of the application of majority voting to pattern recognition,'' in \emph{Proceedings of the 12th IAPR International Conference on Pattern Recognition, Vol. 3 - Conference C: Signal Processing (Cat. No.94CH3440-5)}, vol.~2, 1994, pp. 418--420 vol.2.

\bibitem{BlockchainOracleAdler}
K.~Nelaturu, J.~Adler, M.~Merlini, R.~Berryhill, N.~Veira, Z.~Poulos, and A.~Veneris, ``On public crowdsource-based mechanisms for a decentralized blockchain oracle,'' \emph{IEEE Transactions on Engineering Management}, vol.~67, no.~4, pp. 1444--1458, 2020.

\bibitem{ourpaper}
S.~Aeeneh, N.~Zlatanov, and J.~Yu, ``New bounds on the accuracy of majority voting for multiclass classification,'' \emph{IEEE Transactions on Neural Networks and Learning Systems}, vol.~36, no.~4, pp. 6014--6028, 2025.

\bibitem{truthful}
Y.~Cai, N.~Irtija, E.~E. Tsiropoulou, and A.~Veneris, ``Truthful decentralized blockchain oracles,'' \emph{International Journal of Network Management}, vol.~32, no.~2, p. e2179, 2022.

\bibitem{2021chainlink}
L.~Breidenbach, C.~Cachin, B.~Chan, A.~Coventry, S.~Ellis, A.~Juels, F.~Koushanfar, A.~Miller, B.~Magauran, D.~Moroz \emph{et~al.}, ``Chainlink 2.0: Next steps in the evolution of decentralized oracle networks,'' 2021.

\bibitem{OracleProblem}
\BIBentryALTinterwordspacing
G.~Caldarelli and J.~Ellul, ``The blockchain oracle problem in decentralized finance—a multivocal approach,'' \emph{Applied Sciences}, vol.~11, no.~16, 2021. [Online]. Available: \url{https://www.mdpi.com/2076-3417/11/16/7572}
\BIBentrySTDinterwordspacing

\bibitem{SurveyOracle}
\BIBentryALTinterwordspacing
A.~Pasdar, Y.~C. Lee, and Z.~Dong, ``Connect api with blockchain: A survey on blockchain oracle implementation,'' \emph{ACM Comput. Surv.}, vol.~55, no.~10, feb 2023. [Online]. Available: \url{https://doi.org/10.1145/3567582}
\BIBentrySTDinterwordspacing

\bibitem{9154123}
M.~Kumar, N.~Nikhil, and R.~Singh, ``Decentralising finance using decentralised blockchain oracles,'' in \emph{2020 International Conference for Emerging Technology (INCET)}, 2020, pp. 1--4.

\bibitem{lin2023blockchain}
Y.~Lin, Z.~Gao, H.~Du, D.~Niyato, J.~Kang, Y.~Gao, J.~Wang, and A.~Jamalipour, ``Blockchain-based semantic information sharing and pricing for web 3.0,'' \emph{IEEE Transactions on Network Science and Engineering}, 2023.

\bibitem{yin2020blockchain}
H.~Yin, Z.~Zhang, L.~Zhu, M.~Li, X.~Du, M.~Guizani, and B.~Khoussainov, ``A blockchain-based storage system with financial incentives for load-balancing,'' \emph{IEEE Transactions on Network Science and Engineering}, vol.~8, no.~2, pp. 1178--1188, 2020.

\bibitem{vakilinia2023incentive}
I.~Vakilinia, W.~Wang, and J.~Xin, ``An incentive-compatible mechanism for decentralized storage network,'' \emph{IEEE Transactions on Network Science and Engineering}, 2023.

\bibitem{9321132}
Z.~Peng, J.~Xu, X.~Chu, S.~Gao, Y.~Yao, R.~Gu, and Y.~Tang, ``Vfchain: Enabling verifiable and auditable federated learning via blockchain systems,'' \emph{IEEE Transactions on Network Science and Engineering}, vol.~9, no.~1, pp. 173--186, 2022.

\bibitem{hu2020blockchain}
J.~Hu, K.~Yang, K.~Wang, and K.~Zhang, ``A blockchain-based reward mechanism for mobile crowdsensing,'' \emph{IEEE Transactions on Computational Social Systems}, vol.~7, no.~1, pp. 178--191, 2020.

\bibitem{dataSet}
\BIBentryALTinterwordspacing
M.~VENANZI, W.~Teacy, A.~Rogers, and N.~Jennings, ``Weather sentiment - amazon mechanical turk dataset,'' 2015. [Online]. Available: \url{https://eprints.soton.ac.uk/376543/}
\BIBentrySTDinterwordspacing

\end{thebibliography}

\newpage

\begin{IEEEbiography}[{\includegraphics[width=1in,height=1.25in,clip,keepaspectratio]{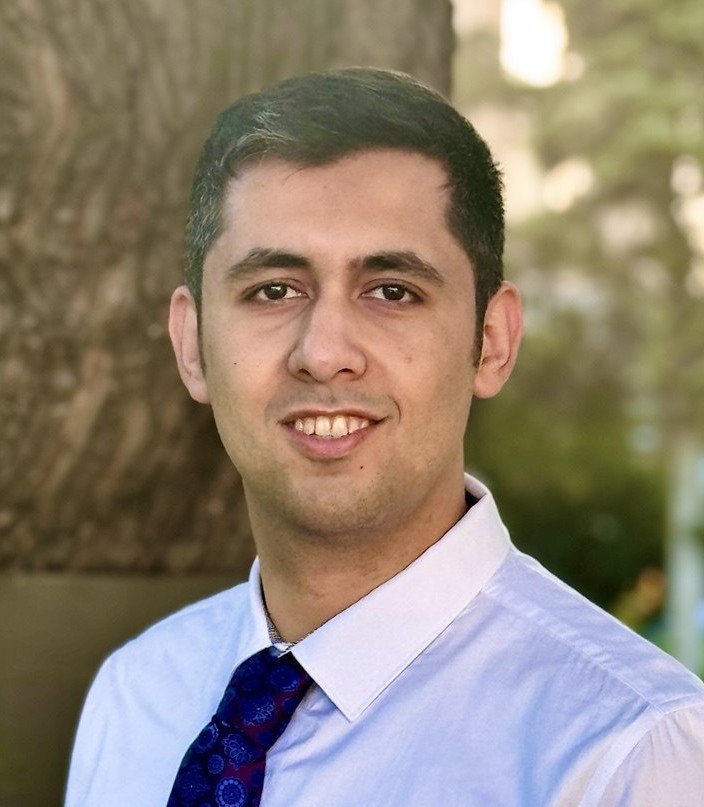}}]{Mohammadsina (Sina) Aeeneh}
Sina Aeeneh is currently pursuing his PhD in the Electrical and Computer Systems Engineering Department at Monash University in Melbourne, Australia. Before joining Monash University, he completed his Bachelor's and Master's degrees at Amirkabir University of Technology and Sharif University of Technology in Tehran, Iran, in 2016 and 2019, respectively. His research interests lie in employing theoretical techniques to analyze electrical and communications systems, with a recent focus on blockchains, cybersecurity, and machine learning. 
\end{IEEEbiography}

\vspace{11pt}

\begin{IEEEbiography}[{\includegraphics[width=1in,height=1.25in,clip,keepaspectratio]{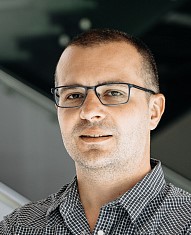}}]{Nikola Zlatanov}
Nikola Zlatanov (Member, IEEE) was born in Macedonia. He received the Dipl.Ing. and master’s degrees in electrical engineering from Ss. Cyril and Methodius University, Skopje, Macedonia, in 2007 and 2010, respectively, and the Ph.D. degree from The University of British Columbia (UBC), Vancouver, BC, Canada, in 2015. Immediately after the Ph.D. degree, he became a Lecturer (U.S. equivalent to an Assistant Professor) with the Department of Electrical and Computer Systems Engineering, Monash University, Melbourne, Australia, where he was promoted to a Senior Lecturer in 2020. Since 2022, he has been a Professor with Innopolis University, Russia. His current research interests include machine learning and wireless communications. He received several scholarships/awards/grants for the work, including the UBC’s Four Year Doctoral Fellowship, in 2010, the UBC’s Killam Doctoral Scholarship and Macedonia’s Young Scientist of the Year, in 2011, the Vanier Canada Graduate Scholarship, in 2012, the Best Journal Paper Award from the German Information Technology Society (ITG), in 2014, the Best Conference Paper Award from ICNC, in 2016, and the ARC Discovery Early Career Researcher Award (DECRA), in 2018. He has served as an Editor for IEEE Communications Letters, from 2015 to 2018, and IEEE Wireless Communications Letters, from 2020 to 2023.\end{IEEEbiography}

\vspace{11pt}

\begin{IEEEbiography}[{\includegraphics[width=1in,height=1.25in,clip,keepaspectratio]{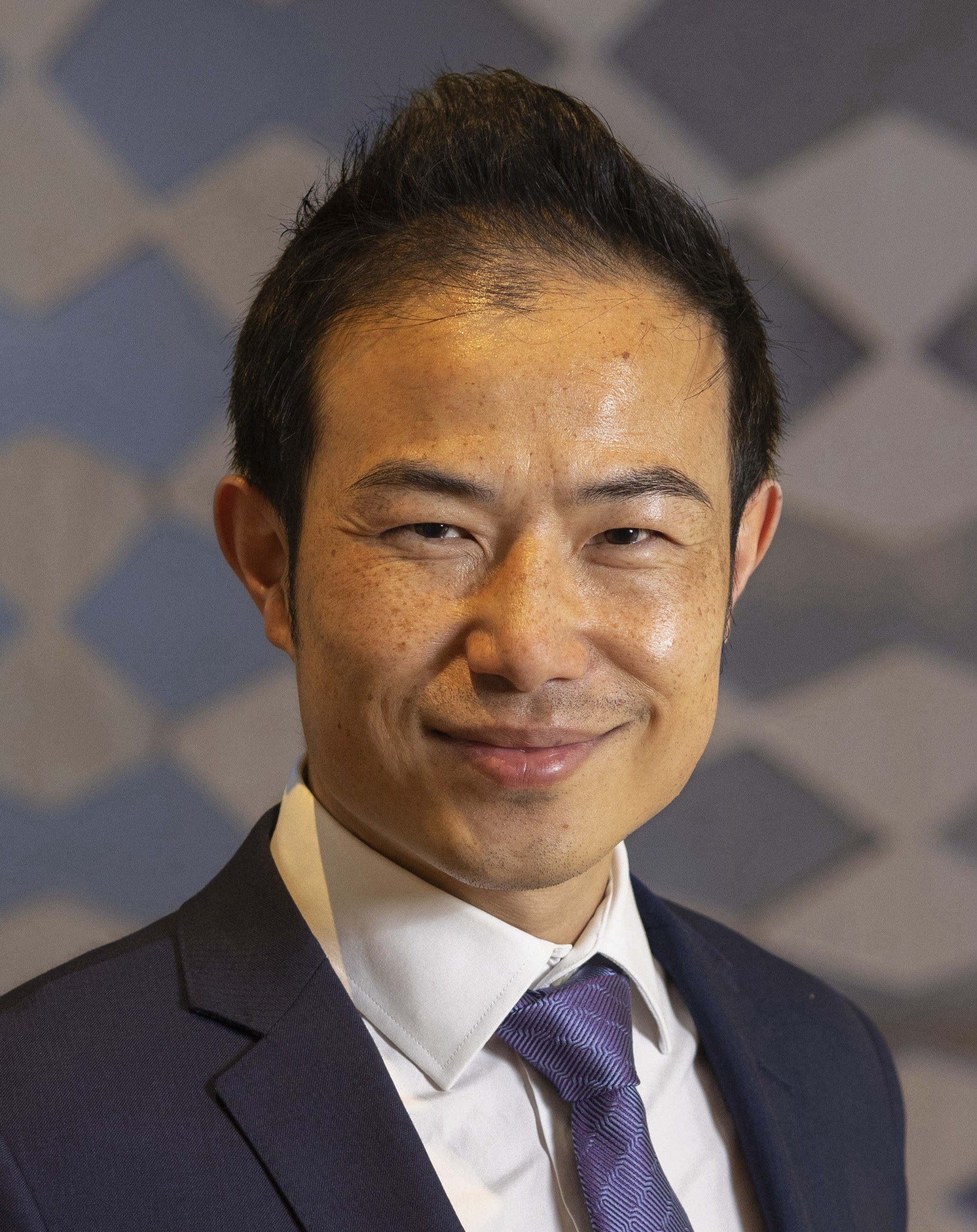}}]{Jiangshan Yu}
Dr. Jiangshan Yu is an Associate Professor at the University of Sydney. He is an elected member of the prestigious IFIP 10.4 Working Group on Dependable Computing and Fault Tolerance and serves on the Scientific Advisory Board for the Austrian Blockchain Center (Austria). Previously, he held positions as Lecturer and Senior Lecturer at Monash University and served as Research Director at Monash Blockchain Technology Centre. Dr. Yu's research interests span trust, reliability, and security, with a particular emphasis on blockchain and trustworthy systems. His work has been widely published in prestigious journals and top-tier conferences such as S\&P, EuroSys, ICDE, DSN, ICDCS, FC, TDSC, and TPDS. Notably, his research has identified critical vulnerabilities in deployed blockchains, leading to recommended fixes adopted by projects totaling over \$30 billion in market capitalization. This impact has garnered widespread media coverage across hundreds of news reports. He has received several competitive awards, including Monash Research Talent Accelerator Award (2023), ARC DECRA (2021), IBM Academic Award (2020). He actively contributes to the academic community as Associate Editor for ACM Distributed Ledger Technologies (DLT), and as a Program Committee member for esteemed conferences such as CCS, VLDB, USENIX ATC, DSN, ICDCS, FC, AFT, ICBC, and Blockchain.
\end{IEEEbiography}

\vfill

\end{document}